\documentclass[onecolumn,aps,prd,groupedaddress,showpacs]{revtex4}%
\usepackage{epsfig}
\usepackage{graphicx}
\usepackage{slashed}
\usepackage{amsmath}
\usepackage{bm}
\usepackage{relsize}
\RequirePackage{xspace}
\begin{document}


\title{\mbox{}\\[10pt]
Relativistic Correction to \mbox{\boldmath $J/\psi$} Production at
Hadron Colliders}

\author{Ying \surname{Fan}}
\email{ying.physics.fan@gmail.com}
\author{Yan-Qing Ma}
\email{yqma.cn@gmail.com}
\author{Kuang-Ta Chao}
\email{ktchao@th.phy.pku.edu.cn}
\affiliation{Department of Physics
and State Key Laboratory of Nuclear Physics and Technology, Peking
University,
 Beijing 100871, China}




\begin{abstract}
Relativistic corrections to the color-singlet $J/\psi$
hadroproduction at the Tevatron and LHC are calculated up to
$\mathcal{O}(v^2)$ in nonrelativistic QCD (NRQCD). The short
distance coefficients are obtained by matching full QCD with NRQCD
results for the subprocess $g+g\to J/\psi+g$. The long distance
matrix elements are extracted from observed $J/\psi$ hadronic and
leptonic decay widths up to $\mathcal{O}(v^2)$. Using the CTEQ6
parton distribution functions, we calculate the LO production cross
sections and relativistic corrections for the process
$p+\bar{p}(p)\rightarrow J/\psi+X$ at the Tevatron and LHC. We find
that the enhancement of $\mathcal{O}(v^2)$ relativistic corrections
to the cross sections over a wide range of large transverse momentum
$p_t$ is negligible, only at a level of about $1\%$. This tiny
effect is due to the smallness of the correction to short distance
coefficients and the suppression from long distance matrix elements.
These results indicate that relativistic corrections can not help to
resolve the large discrepancy between leading order prediction and
experimental data for $J/\psi$ production at the Tevatron.
\end{abstract}
\pacs{12.38.Bx,12.39.St,13.85.Ni,14.40.Gx}

\maketitle


\section{Introduction}
Nonrelativistic QCD (NRQCD)\cite{Bodwin:1994jh} is an effective
field theory to describe production and decay of heavy quarkonium.
In this formalism, inclusive production cross sections and decay
widths of charmonium and bottomnium can be factored into short
distance coefficients, indicating the creation or annihilation of a
heavy quark pair, and long distance matrix elements, representing
the evolvement of a free quark pair into a bound state. The short
distance part can be calculated perturbatively in powers of coupling
constant $\alpha_s$, while the nonperturbative matrix elements,
which are scaled as $v$, the typical velocity of heavy quark or
antiquark in the meson rest frame, can be estimated by
nonperturbative methods or models, or extracted from experimental
data.

One important aspect of NRQCD is the introduction of color-octet
mechanism, which allows the intermediate heavy quark pair to exist
in a color-octet state at short distances and evolve into the
color-singlet bound state at long distances. This mechanism has been
applied successfully to absorb the infrared divergences in
P-wave\cite{Bodwin:1994jh,Huang,Petrelli:1997} and
D-wave\cite{Heshort,Fan:2008} decay widths of heavy quarkonia. In
Ref. \cite{Braaten:1994}, the color-octet mechanism was introduced
to account for the $J/\psi$ production at the Tevatron, and the
theoretical prediction of production rate fits well with
experimental data. However, the color-octet gluon fragmentation
predicts that the $J/\psi$ is transversely polarized at large
transverse momentum $p_t$, which is in contradiction with the
experimental data\cite{CDF:2007}. (For a review of these issues, one
can refer to
Refs.\cite{Kramer:2001,Brambilla:2004wf,Lansberg:2006dh}). Moreover,
in Refs.\cite{Mayq:2008,Gong:2009} it was pointed out that the
color-octet long distance matrix elements of $J/\psi$ production may
be much smaller than previously expected, and accordingly this may
reduce the color-octet contributions to $J/\psi$ production at the
Tevatron.

In the past a couple of years, in order to resolve the large
discrepancy between the color-singlet leading-order (LO) predictions
and experimental
measurements\cite{CDF:1992,CDF:1997-572,CDF:1997-578} of $J/\psi$
production at the Tevatron, the next-to-leading-order (NLO) QCD
corrections to this process have been performed, and a large
enhancement of an order of magnitude for the cross section at large
$p_t$ is found\cite{Campbell:2007,Gong:2008}. But this still can not
make up the large discrepancy between the color-singlet contribution
and data. Similarly, the observed double charmonium production cross
sections in $e^+e^-$ annihilation at B
factories\cite{Abe:2002rb,BaBar:2005} also significantly differ from
LO theoretical predictions\cite{Braaten:2002fi}. Much work has been
done and it seems that those discrepancies could be resolved by
including NLO QCD
corrections\cite{Zhang:2005cha,Zhang:2006ay,Gong:2007db,zhangma08}
and relativistic corrections\cite{bodwin06,He:2007te}. One may
wonder if the relativistic correction could also play a role to some
extent in resolving the long standing puzzle of $J/\psi$ production
at the Tevatron.

In this paper we will estimate the effect of relativistic
corrections to the color-singlet $J/\psi$ production based on NRQCD.
The relativistic effects are characterized by the relative velocity
$v$ with which the heavy quark or antiquark moves in the quarkonium
rest frame. According to the velocity scaling rules of
NRQCD\cite{velocity}, the matrix elements of operators can be
organized into a hierarchy in the small parameter $v$.
We calculate the short distance part perturbatively up to
$\mathcal{O}(v^2)$. In order to avoid model dependence in
determining the long distance matrix elements, we extract the matrix
elements of up to dimension-8 four fermion operators from observed
decay rates of $J/\psi$\cite{Amsler:2008zz}. We find that the
relativistic effect on the color-singlet $J/\psi$ production at both
the Tevatron and LHC is tiny and negligible, and relativistic
corrections can not offer much help to resolve the puzzle associated
with charmonium production at the Tevatron, and other mechanisms
should be investigated to clarify the problem.

The rest of the paper is organized as follows. In Section II, the
NRQCD factorization formalism and matching condition of full QCD and
NRQCD effective field theory at long distances are described
briefly, and then detailed calculations are given, including the
perturbative calculation of short-distance coefficient, the
long-distance matrix elements extracted from experimental data, and
the parton level differential cross section convolution with the
parton distribution functions (PDF). In Section III, numerical
results of differential cross sections over transverse momentum
$p_t$ at the Tevatron and LHC are given and discussions are made for
the enhancement effects of relativistic corrections. Finally the
summary of this work is presented.

\section{Production Cross Section in NRQCD Factorization}
According to NRQCD factorization\cite{Bodwin:1994jh}, the inclusive
cross section for the hadroproduction of $J/\psi$ can be written
as
\begin{equation}
\label{factorization} \frac{d\sigma}{dt}(g+g \rightarrow
J/\psi+g)=\sum_n\frac{F_n}{m_c^{d_n-4}}\langle0|\mathcal{O}_n^{J/\psi}|0\rangle.
\end{equation}
The short-distance coefficients $F_n$ describe the production of a
heavy quark pair $\mathcal{Q}\bar{\mathcal{Q}}$ from the gluons,
which come from the initial state hadrons, and usually expressed in
kinematic invariants. $m_c$ is the mass of charm quark. The
long-distance matrix elements
$\langle0|\mathcal{O}_n^{J/\psi}|0\rangle$ with mass dimension $d_n$
describe the evolution of $\mathcal{Q}\bar{\mathcal{Q}}$ into
$J/\psi$. The subscript $n$ represents the configuration in which
the $c\bar c$ pair can be for the $J/\psi$ Fock state expansion, and
it is usually denoted as $n={}^{2S+1}L_{J}^{[1,8]}$. Here, $S$,$L$
and $J$ stand for spin, orbital and total angular momentum of the
heavy quarkonium respectively. Superscript 1 or 8 means the
color-singlet or color-octet state.

For the color-singlet ${}^3S_1$ $c\bar c$ production, there are only
two matrix elements contributing up to $\mathcal{O}(v^2)$: the
leading order term
$\langle0|\mathcal{O}^{J/\psi}({}^3S_1^{[1]})|0\rangle$ and the
relativistic correction term
$\langle0|\mathcal{P}^{J/\psi}({}^3S_1^{[1]})|0\rangle$. Therefore
the differential cross section takes the following form
\begin{eqnarray}\label{crosssection}
\frac{d\sigma}{dt}(g+g \rightarrow
J/\psi+g)&=&\frac{F({}^3S_1^{[1]})}{m_c^2}\langle0|\mathcal{O}^{J/\psi}({}^3S_1^{[1]})|0\rangle
+\frac{G({}^3S_1^{[1]})}{m_c^4}\langle0|\mathcal{P}^{J/\psi}({}^3S_1^{[1]})|0\rangle+\mathcal{O}(v^4).
\end{eqnarray}
And the explicit expressions of the matrix elements
are\cite{Bodwin:1994jh}:
\begin{eqnarray}\label{matrixelement}
\langle0|\mathcal{O}^{J/\psi}({}^3S_1^{[1]})|0\rangle&=&\langle0|\chi^\dagger\sigma^{i}\psi(a^\dagger_{\psi}a_{\psi})
\psi^\dagger\sigma^i\chi|0\rangle,
\nonumber\\
\langle0|\mathcal{P}^{J/\psi}({}^3S_1^{[1]})|0\rangle&=&\langle0|\frac{1}{2}[\chi^\dagger\sigma^{i}\psi
(a^\dagger_{\psi}a_{\psi})\psi^\dagger\sigma^i(-\frac{i}{2}\overleftrightarrow{\mathbf{D}})^2\chi+h.c.]|0\rangle,
\end{eqnarray}
where $\psi$ annihilates a heavy quark, $\chi$ creates a heavy
antiquark, $a^\dagger_{\psi}$ and $a_{\psi}$ are operators creating
and annihilating $J/\psi$ in the final state, and
$\overleftrightarrow{\mathbf{D}}=\overrightarrow{\mathbf{D}}-\overleftarrow{\mathbf{D}}$.

In order to determine the short distance coefficients
$F({}^3S_1^{[1]})$ and $G({}^3S_1^{[1]})$, matching condition of
full QCD and NRQCD is needed:
\begin{equation}\label{matching}
\frac{d\sigma}{dt}(g+g \rightarrow
J/\psi+g)\Big|_{pert~QCD}=\frac{F({}^3S_1^{[1]})}{m_c^2}\langle0|\mathcal{O}^{J/\psi}({}^3S_1^{[1]})|0\rangle
+\frac{G({}^3S_1^{[1]})}{m_c^4}\langle0|\mathcal{P}^{J/\psi}({}^3S_1^{[1]})|0\rangle\Big|_{pert~NRQCD}.
\end{equation}

The differential cross section for the production of charmonium
$J/\psi$ on the left hand side of Eq.(\ref{matching}) can be
calculated in perturbative QCD. On the right hand side the long
distance matrix elements are extracted from experimental data.
Quantities on both sides of the equation are expanded at leading
order of $\alpha_s$ and next-to-leading order of $v^2$. Then the
short distance coefficients $F({}^3S_1^{[1]})$ and
$G({}^3S_1^{[1]})$ can be obtained by comparing the terms with
powers of $v^2$ on both sides.

\subsection{Perturbative Short Distance Coefficients}
\begin{figure}
\begin{center}
\includegraphics{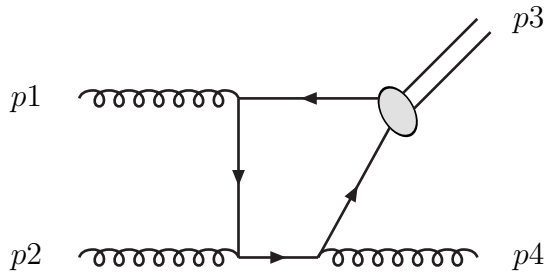}
\caption{\label{fig:ggtojg}Typical Feynman diagram for
${}^3S_1^{[1]}$ $c\bar c$ hadroproduction at LO. }
\end{center}
\end{figure}

We now present the calculation of relativistic correction to the
process $g+g\rightarrow J/\psi+g$. In order to determine the
$\mathcal{O}(v^2)$ contribution in Eq.(\ref{crosssection}), the
differential cross section on the left hand side of
Eq.(\ref{matching}) or equivalently the QCD amplitude should be
expanded up to $\mathcal{O}(v^2)$. 
We use {\tt FeynArts}~\cite{feynarts} to generate Feynman diagrams
and amplitudes, {\tt FeynCalc}~\cite{Mertig:an} to handle
amplitudes, and Fortran to evaluate the phase space integrations. A
typical Feynman diagram for the process is shown in Fig.
\ref{fig:ggtojg}.
The momenta of quark and antiquark in the lab frame
are\cite{He:2007te,Ma:2000,fourmomenta}:
\begin{eqnarray}
\frac{1}{2}P+q&=&L(\frac{1}{2}P_r+q_r),\nonumber\\
\frac{1}{2}P-q&=&L(\frac{1}{2}P_r-q_r),
\end{eqnarray}
where $L$ is the Lorenz boost matrix from the rest frame of the
$J/\psi$ to the frame in which it moves with four momentum $P$.
$P_r=(2E_q,\mathbf{0})$, $E_q=\sqrt{m_c^2+|\vec{q}|^2}$, and
$2q_r=2(0,\vec{q})$ is the relative momentum between heavy quark and
antiquark in the $J/\psi$ rest frame. The differential cross section
on the left hand side of Eq.~(\ref{matching}) is
\begin{equation}\label{perQCD}
\frac{d\sigma}{dt}(g+g \rightarrow
J/\psi+g)\Big|_{pert~QCD}=\frac{1}{16\pi
s^2}\overline{\sum}|\mathcal{M}(g+g \rightarrow
J/\psi+g)|^2\langle0|\mathcal{O}^{J/\psi}({}^3S_1^{[1]})|0\rangle,
\end{equation}
where $\langle0|\mathcal{O}^{J/\psi}({}^3S_1^{[1]})|0\rangle$ is the
matrix element evaluated at tree level, and summation/average of the
color and polarization degrees of freedom for the final/initial
state has been implied by the symbol $\overline{\sum}$. The
amplitude for the color-singlet process $g(p_1)+g(p_2)\rightarrow
J/\psi(p_3=P)+g(p_4)$ is
\begin{equation}\label{amplitude}
\mathcal{M}(g+g \rightarrow
J/\psi+g)=\sqrt{\frac{1}{E_q}}Tr[\mathcal{C}^{[1]}\Pi^{(1)}\mathcal{M}^{am}],
\end{equation}
where $\mathcal{M}^{am}$ denotes the parton-level amplitude
amputated of the heavy quark spinors. The factor
$\sqrt{\frac{1}{E_q}}$ comes from the normalization of the composite
state $|{}^3S_1^{[1]}\rangle$\cite{Fan:2008}. Here the covariant
projection operator method\cite{Bodwin:2002,Keung:1983} is adopted.
For a color-singlet state, the color projector
$\mathcal{C}^{[1]}=\frac{\delta_{ij}}{\sqrt{N_c}}$. The covariant
spin-triplet projector $\Pi^{(1)}$ in (\ref{amplitude}) is defined
by
\begin{equation}
\Pi^{(1)}=\sum_{s\bar{s}}v(s)\bar{u}(\bar{s})\langle\frac{1}{2},s;\frac{1}{2},\bar{s}|1,S_z\rangle,
\end{equation}
with its explicit form
\begin{equation}\label{projector}
\Pi^{(1)}=\frac{1}{\sqrt{2}(E_q+m_c)}(\frac{\slashed{P}}{2}-\slashed{q}-m_c)\frac{\slashed{P}-2E_q}{4E_q}
\slashed{\epsilon}(\frac{\slashed{P}}{2}+\slashed{q}+m_c),
\end{equation}
where the superscript $(1)$ denotes the spin triplet state and
$\epsilon_{\rho}$ is the polarization vector of the spin 1 meson.
The Lorentz-invariant Mandelstam variables are defined by
\begin{eqnarray}
s&=&(p_1+p_2)^2=(p_3+p_4)^2,\nonumber\\
t&=&(p_1-p_3)^2=(p_2-p_4)^2,\nonumber\\
u&=&(p_1-p_4)^2=(p_2-p_3)^2,
\end{eqnarray}
and they satisfy
\begin{equation}\label{stu}
s+t+u=P^2=4E_q^2=4(m_c^2+|\vec{q}|^2).
\end{equation}
Furthermore, the covariant spinors are normalized relativistically
as $\bar{u}u=-\bar{v}v=2m_c$.

Let $\mathcal{M}$ be short for the amplitude $\mathcal{M}(g+g
\rightarrow J/\psi+g)$ in Eq.~(\ref{amplitude}), and it can be
expanded in powers of $v$ or equivalently $|\vec{q}|$. That is
\begin{equation}
\mathcal{M}=\epsilon_{\rho}\mathcal{M}^{\rho}=\epsilon_{\rho}(\mathcal{M}^{\rho}(0)+\frac{1}{2}q^{\alpha}q^{\beta}
\frac{\partial^2\mathcal{M}^{\rho}}{\partial q^{\alpha}\partial
q^{\beta}}\Big|_{q=0})+\mathcal{O}(q^4),
\end{equation}
where high order terms in four momentum $q$ have been omitted. Terms
of odd powers in $q$ vanish because the heavy quark pair is in an S
wave configuration. Note that the polarization vector
$\epsilon_{\rho}$ also depends on $q$, but it only has even powers
of four momentum $q$, and their expressions may be found e.g. in the
Appendix of Ref.~\cite{Chang:2004}. Therefore expansion on $q^2$ of
$\epsilon_{\rho}$ can be carried out after amplitude squaring. The
following substitute is adopted:
\begin{equation}
q^{\alpha}q^{\beta}=\frac{1}{3}|\vec{q}|^2(-g^{\alpha\beta}+\frac{P^\alpha
P^\beta}{P^2})\equiv\frac{1}{3}|\vec{q}|^2\Delta^{\alpha\beta}.
\end{equation}
This substitute should be understood to hold in the integration over
relative momentum $\vec{q}$ and in the S wave case. Here,
$|\vec{q}|^2$ can be identified as\cite{Bodwin:2007,Bodwin:2002}
\begin{equation}
|\vec{q}|^2=\frac{|\langle0|\chi^\dagger\mbox{\boldmath
$\sigma$}(-\frac{i}{2}\overleftrightarrow{\mathbf{D}})^2\psi|\psi({}^3S_1^{[1]})\rangle|}
{|\langle0|\chi^\dagger\mbox{\boldmath
$\sigma$}\psi|\psi({}^3S_1^{[1]})\rangle|}
=\frac{\langle0|\mathcal{P}^{J/\psi}({}^3S_1^{[1]})|0\rangle}
{\langle0|\mathcal{O}^{J/\psi}({}^3S_1^{[1]})|0\rangle}[1+\mathcal{O}(v^4)].
\end{equation}
Then the amplitude squared defined in Eq.~(\ref{perQCD}) up to
$\mathcal{O}(v^2)$ are:
\begin{equation}\label{ampsquare}
\sum|\mathcal{M}|^2=\mathcal{M}^{\rho}(0)\mathcal{M}^{{\lambda}*}(0)\sum\epsilon_{\rho}\epsilon^*_{\lambda}
+\frac{1}{6}|\vec{q}|^2\left[(\Delta^{\alpha\beta}\frac{\partial^2\mathcal{M}^{\rho}}{\partial
q^{\alpha}\partial q^{\beta}})\Big|_{q=0}\mathcal{M}^{{\lambda}*}(0)
+(\Delta^{\alpha\beta}\frac{\partial^2\mathcal{M}^{{\lambda}*}}{\partial
q^{\alpha}\partial
q^{\beta}})\Big|_{q=0}\mathcal{M}^{\rho}(0)\right]\sum\epsilon_{\rho}\epsilon^*_{\lambda}+\mathcal{O}(v^4).
\end{equation}
The heavy quark and antiquark are taken to be on shell, which means
that $P\cdot q=0$, and then gauge invariance is maintained. The
polarization sum in Eq. (\ref{ampsquare}) is
\begin{equation}
\sum\epsilon_{\rho}\epsilon^*_{\lambda}=-g_{\rho\lambda}+\frac{P_\rho
P_\lambda}{P^2}~.
\end{equation}
It is clearly seen that the polarization sum above only contains
even order powers of four momentum $q$, therefore it will make
contribution to relativistic correction at $\mathcal{O}(v^2)$ in the
first term on the right hand side of Eq.~(\ref{ampsquare}) when
contraction over indices $\rho$ and $\lambda$ is carried out.
However, since the second term on the right hand side of
Eq.~(\ref{ampsquare}) already has term proportional to $q^2$, i.e.
$|\vec{q}|^2$, the four momentum $q$ can be set to zero throughout
the index contraction. Then we have
\begin{equation}
\sum|\mathcal{M}|^2=A+B|\vec{q}|^2+\mathcal{O}(v^4),
\end{equation}
where $A$ and $B$ are independent of $|\vec{q}|$. By comparing
Eq.(\ref{matching}) and Eq.(\ref{perQCD}), we obtain the short
distance coefficients shown explicitly below. The leading order one
is
\begin{eqnarray}\label{F3S1}
\frac{F({}^3S_1^{[1]})}{m_c^2}&=&\frac{1}{16\pi
s^2}\frac{1}{64}\frac{1}{4}\frac{1}{2N_c}\frac{1}{3}~A\nonumber\\
&=&\frac{1}{16\pi
s^2}\frac{1}{64}\frac{1}{4}\frac{1}{2N_c}\frac{1}{3}(4\pi\alpha_s)^3
5120 m_c [16 \left(s^2+t s+t^2\right) m_c^4-4 \left(2 s^3+3 t s^2+3
t^2 s+2 t^3\right) m_c^2\nonumber\\
&&+\left(s^2+t
   s+t^2\right)^2]/[9 \left(s-4 m_c^2\right)^2 \left(t-4 m_c^2\right)^2
   (s+t)^2],
\end{eqnarray}
and the relativistic correction term is
\begin{eqnarray}\label{G3S1}
\frac{G({}^3S_1^{[1]})}{m_c^4}&=&\frac{1}{16\pi
s^2}\frac{1}{64}\frac{1}{4}\frac{1}{2N_c}\frac{1}{3}~B\nonumber\\
&=&\frac{1}{16\pi
s^2}\frac{1}{64}\frac{1}{4}\frac{1}{2N_c}\frac{1}{3}(4\pi\alpha_s)^3(-2560)
[2048 \left(3 s^2+2 t s+3 t^2\right) m_c^{10}-256 \left(5
s^3-2 ts^2-2 t^2 s+5 t^3\right) m_c^8\nonumber\\
&& -320 \left(3s^4+10 t s^3+10 t^2 s^2+10 t^3 s+3 t^4\right)
m_c^6+16 \left(21 s^5+63 t s^4+88 t^2 s^3+88 t^3 s^2+63 t^4 s+21
t^5\right)
   m_c^4\nonumber\\
   &&-4 \left(7 s^6+18 t s^5+23 t^2 s^4+28 t^3 s^3+23 t^4 s^2+18 t^5 s+7 t^6\right)
   m_c^2
  -s t (s+t) \left(s^2+t
   s+t^2\right)^2]/\nonumber\\
   &&[27 m_c \left(4 m_c^2-s\right)^3 \left(4 m_c^2-t\right)^3
   (s+t)^3].
\end{eqnarray}
Each of the factors has its own origin: $1/16\pi s^2$ is
proportional to the inverse square of the M${\o}$ller's invariant
flux factor, $1/64$ and $1/4$ are the color average and spin average
of initial two gluons respectively, $1/2N_c$ comes from the
color-singlet long distance matrix element definition in Eq.
(\ref{matrixelement}) with $N_c$=3, $1/3$ is the spin average for
total spin $J=1$ states and $(4\pi\alpha_s)^3$ quantifies the
coupling in the QCD interaction vertices. Furthermore the variable
$u$ has been expressed in terms of $s$ and $t$ through
Eq.(\ref{stu}). To verify our results, we find that those in
Ref.~\cite{Ma:2000} discussed for $J/\psi$ photoproduction are
consistent with ours under replacement
$(4\pi\alpha)e_c^2\rightarrow(4\pi\alpha_s)$, and the result in
Ref.\cite{Leibovich:1996} agrees with ours at leading order after
performing the polarization summation.

\subsection{Nonperturbative Long Distance Matrix Elements}
The long distance matrix elements may be determined by potential
model\cite{Bodwin:2007,bodwin06} or lattice
calculations\cite{Sinclair:1997}, and by phenomenological extraction
from experimental data. Here we first extract the decay matrix
elements from experimental data. Up to NLO QCD and $v^2$
relativistic corrections, decay widths of the color-singlet $J/\psi$
to light hadrons (LH) and $e^+e^-$ can be expressed analytically as
follows\cite{Bodwin:2002}:
\begin{eqnarray}
\Gamma[J/\psi\rightarrow
LH]&=&\frac{F_{LH}({}^3S_1^{[1]})}{m_c^2}\langle
H|\mathcal{O}^{J/\psi}({}^3S_1^{[1]})|H\rangle+
\frac{G_{LH}({}^3S_1^{[1]})}{m_c^4}\langle H|\mathcal{P}^{J/\psi}({}^3S_1^{[1]})|H\rangle,\nonumber\\
\Gamma[J/\psi\rightarrow
e^+e^-]&=&\frac{F_{e^+e^-}({}^3S_1^{[1]})}{m_c^2}\langle
H|\mathcal{O}^{J/\psi}({}^3S_1^{[1]})|H\rangle+
\frac{G_{e^+e^-}({}^3S_1^{[1]})}{m_c^4}\langle
H|\mathcal{P}^{J/\psi}({}^3S_1^{[1]})|H\rangle,
\end{eqnarray}
where the short distance coefficients are\cite{Bodwin:2002}
\begin{eqnarray}\label{FG}
F_{LH}({}^3S_1^{[1]})&=&\frac{(N_c^2-1)(N_c^2-4)}{N_c^3}\frac{(\pi^2-9)}{18}\alpha_s^3(2m_c)
[1+(-9.46C_F+4.13C_A-1.161N_f)\frac{\alpha_s}{\pi}]\nonumber\\
&&+2\pi e_Q^2(\sum_{i=1}^{Nf}
Q_i^2)\alpha_e^2(1-\frac{13}{4}C_F\frac{\alpha_s}{\pi}),\nonumber\\
G_{LH}({}^3S_1^{[1]})&=&-\frac{5(19\pi^2-132)}{729}\alpha_s^3(2m_c),\nonumber\\
F_{e^+e^-}({}^3S_1^{[1]})&=&\frac{2\pi
e_Q^2\alpha_e^2}{3}[1-4C_F\frac{\alpha_s(2m_c)}{\pi}],\nonumber\\
G_{e^+e^-}({}^3S_1^{[1]})&=&-\frac{8\pi e_Q^2\alpha_e^2}{9}.
\end{eqnarray}
Then, the production matrix elements can be related to the decay
matrix elements through vacuum saturation approximation
\begin{equation}\label{2J+1}
\langle 0|\mathcal{O}^{J/\psi}({}^3S_1^{[1]})|0\rangle=(2J+1)\langle
H|\mathcal{O}^{J/\psi}({}^3S_1^{[1]})|H\rangle[1+\mathcal{O}(v^4)]=3\langle
H|\mathcal{O}^{J/\psi}({}^3S_1^{[1]})|H\rangle[1+\mathcal{O}(v^4)].
\end{equation}
Combining the above equations and the experimental data in
\cite{Amsler:2008zz}, i.e., $\Gamma[J/\psi\rightarrow LH]=81.7$~KeV
and $\Gamma[J/\psi\rightarrow e^+e^-]=5.55$~KeV and excluding the
NLO QCD radiative corrections in (\ref{FG}), we get the solutions
accurate at leading order in $\alpha_s$
\begin{equation}\label{matrixlo}
\langle
0|\mathcal{O}^{J/\psi}({}^3S_1^{[1]})|0\rangle=0.868GeV^3,~~~~~~~\langle
0|\mathcal{P}^{J/\psi}({}^3S_1^{[1]})|0\rangle=0.190GeV^5.
\end{equation}
And the enhanced matrix elements accurate up to NLO in $\alpha_s$
can be obtained by including NLO QCD radiative corrections in
(\ref{FG})
\begin{equation}\label{matrixenhanced}
\langle
0|\mathcal{O}^{J/\psi}({}^3S_1^{[1]})|0\rangle=1.64GeV^3,~~~~~~~\langle
0|\mathcal{P}^{J/\psi}({}^3S_1^{[1]})|0\rangle=0.320GeV^5.
\end{equation}
The strong coupling constant evaluated at the charm quark mass scale
is $\alpha_s(2m_c)=0.250$ for $m_c=1.5$GeV. The other input
parameters are chosen as follows: the QCD scale parameter
$\Lambda_{QCD}=392$MeV, the number of quarks with mass less than the
energy scale $m_c$ is $N_f=3$, color factor $C_F=4/3$ and $C_A=3$,
the electric charge of charm quark is $e_Q=2/3$, $Q_i$ are the
electric charges of the light quarks and fine structure constant
$\alpha_e=1/137$. Our numerical values for the production matrix
elements $\langle 0|\mathcal{O}^{J/\psi}({}^3S_1^{[1]})|0\rangle$
and $\langle 0|\mathcal{P}^{J/\psi}({}^3S_1^{[1]})|0\rangle$ are
accurate up to NLO in $v^2$ with uncertainties due to experimental
errors and higher order corrections.

\subsection{Cross Sections for $p+\bar{p}(p)\rightarrow J/\psi+X$ and Phase Space Integration}
Based on the results obtained for the subprocess $g+g\to J/\psi+g$
we further calculate the production cross sections and relativistic
corrections in the process $p+\bar{p}(p)\rightarrow J/\psi+X$, which
involves hadrons as the initial states. In order to get the cross
sections at the hadron level, the partonic cross section defined in
Eq.(\ref{perQCD}) has to be convoluted with the parton distribution
function (PDF) $f_{g/p}(x_{\pm})$, where $x_{\pm}$ denotes the
fraction of the proton or antiproton beam energy carried by the
gluons.

We will work in the $p\bar{p}$ center-of-mass (CM) frame and denote
the $p\bar{p}$ energy by $\sqrt{S}$, the rapidity of $J/\psi$ by
$y_C$, and that of the gluon jet by $y_D$. The differential cross
section of $p+\bar{p}(p)\rightarrow J/\psi+X$ can be written
as\cite{klasen:2001}
\begin{equation}
\frac{d^3\sigma(p+\bar{p}(p)\rightarrow
J/\psi+X)}{dp_t^2dy_Cdy_D}=x_{+}f_{g/p}(x_{+})x_{-}f_{g/\bar{p}(p)}(x_{-})\frac{d\sigma(g+g\rightarrow
J/\psi+g)}{dt},
\end{equation}
where
\begin{equation}
x_{\pm}=\frac{m_t^Cexp(\pm y_C)+m_t^Dexp(\pm y_D)}{\sqrt{S}},
\end{equation}
with the transverse mass $m_t^{C,D}=\sqrt{m^2_{C,D}+p^2_t}$, the
meson mass $m_C=2m_c$, the gluon mass $m_D=0$ and the transverse
momentum $p_t$. The Mandelstam variables can be expressed in terms
of $p_t$, $y_C$ and $y_D$
\begin{eqnarray}
s&=&x_{+}x_{-}S,\nonumber\\
t&=&-p_t^2-m_t^{C}m_t^{D}exp(y_D-y_C),\nonumber\\
u&=&-p_t^2-m_t^{C}m_t^{D}exp(y_C-y_D).
\end{eqnarray}
The accessible phase space puts kinetic constraints on variables
$p_t$, $y_C$ and $y_D$ for a fixed value of two colliding hadron
center-of-mass energy $\sqrt{S}$
\begin{eqnarray}
0&\leq& p_t \leq
\frac{1}{2}\sqrt{\frac{\lambda(S,m_C^2,m_D^2)}{S}},\nonumber\\
|y_C|&\leq& Arcosh\frac{S+m_C^2-m_D^2}{2\sqrt{S}m_t^C},\nonumber\\
-\ln\frac{\sqrt{S}-m_t^Cexp(-y_C)}{m_t^D}&\leq&
y_D\leq\ln\frac{\sqrt{S}-m_t^Cexp(y_C)}{m_t^D},
\end{eqnarray}
where $\lambda(x,y,z)=x^2+y^2+z^2-2(xy+yz+zx)$. The distribution
over $p_t$ of differential cross section can be obtained after phase
space integration.

\section{Numerical Results and Analysis}
\begin{figure}
\begin{center}
\includegraphics[bb=10 20 300 200]{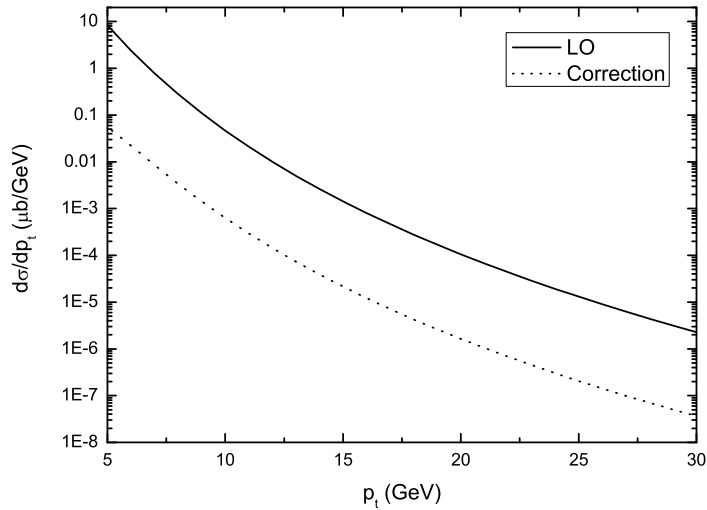}
\caption{\label{fig:Teva}The $p_t$ distribution of
$d\sigma(p+\bar{p}\rightarrow J/\psi+X)/dp_t$ at the Tevatron with
$\sqrt{S}=1.96~TeV$. The $\mathcal{O}(v^0)$ and $\mathcal{O}(v^2)$
results are represented by the solid and dotted lines respectively.}
\end{center}
\end{figure}

\begin{figure}
\begin{center}
\includegraphics[bb=10 20 300 200]{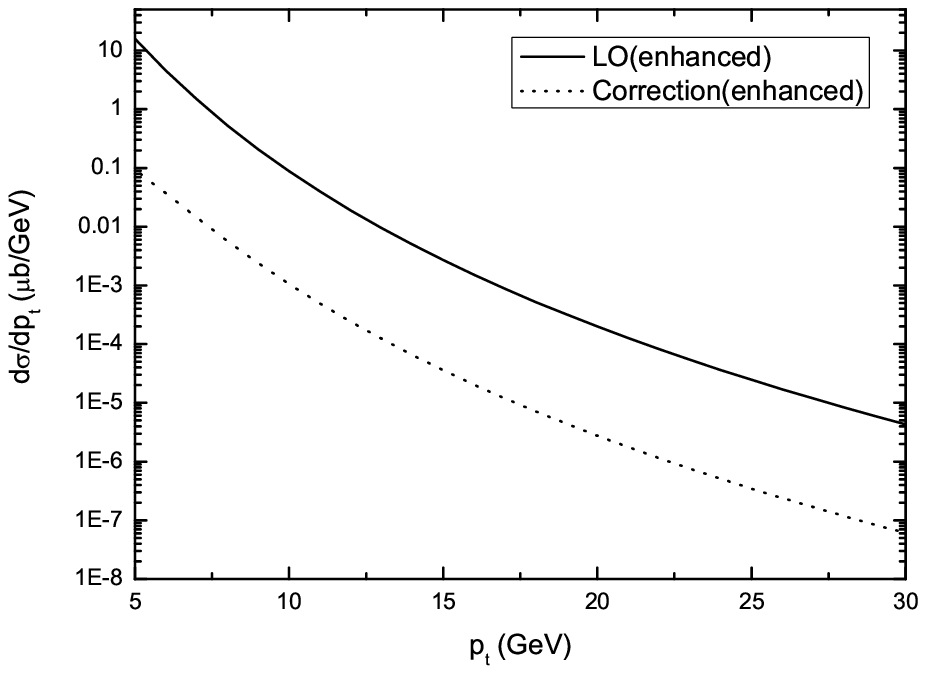}
\caption{\label{fig:Teva(enhanced)}The $p_t$ distribution of
$d\sigma(p+\bar{p}\rightarrow J/\psi+X)/dp_t$ (with enhanced matrix
elements) at the Tevatron with $\sqrt{S}=1.96~TeV$. The
$\mathcal{O}(v^0)$ and $\mathcal{O}(v^2)$ results are represented by
the solid and dotted lines respectively.}
\end{center}
\end{figure}

\begin{figure}
\begin{center}
\includegraphics[bb=10 20 300 200]{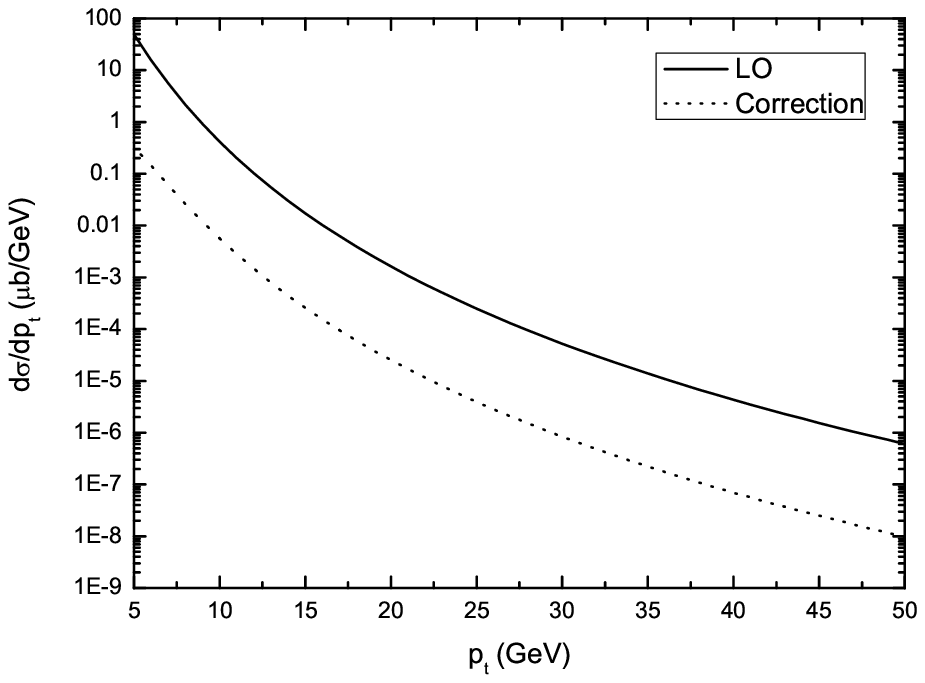}
\caption{\label{fig:LHC14TeV}The $p_t$ distribution of
$d\sigma(p+p\rightarrow J/\psi+X)/dp_t$ at the LHC with
$\sqrt{S}=14~ TeV$. The $\mathcal{O}(v^0)$ and $\mathcal{O}(v^2)$
results are represented by the solid and dotted lines respectively.}
\end{center}
\end{figure}

\begin{figure}
\begin{center}
\includegraphics[bb=10 20 300 200]{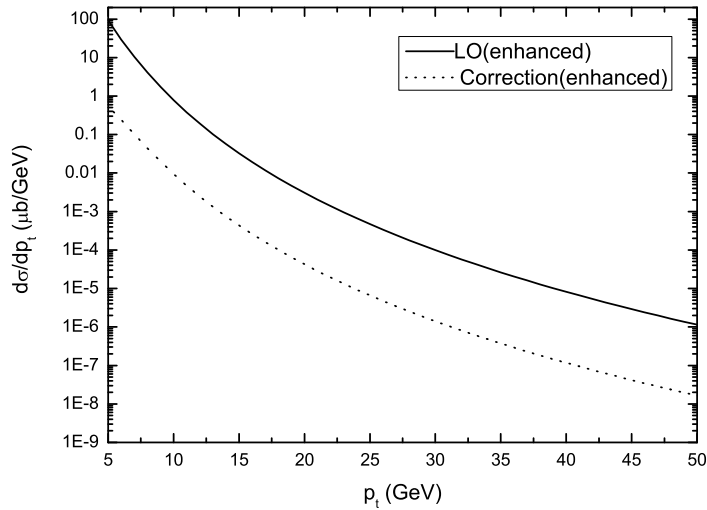}
\caption{\label{fig:LHC14TeV(enhanced)}The $p_t$ distribution of
$d\sigma(p+p\rightarrow J/\psi+X)/dp_t$ (with enhanced matrix
elements) at the LHC with $\sqrt{S}=14~TeV$. The $\mathcal{O}(v^0)$
and $\mathcal{O}(v^2)$ results are represented by the solid and
dotted lines respectively.}
\end{center}
\end{figure}

The CTEQ6 PDFs\cite{Pumplin:2002} are used in our numerical
calculation. We present the distribution of $J/\psi$ production
differential cross section $d\sigma/dp_t$ over $p_t$ at the Tevatron
with $\sqrt{S}=1.96$~TeV and at the LHC with $\sqrt{S}=14$~TeV in
Fig.\ref{fig:Teva}, Fig.\ref{fig:Teva(enhanced)} and
Fig.\ref{fig:LHC14TeV}, Fig.\ref{fig:LHC14TeV(enhanced)}. 
The solid line represents the distribution at leading order in
$\mathcal{O}(v^2)$, and the dotted line describes the relativistic
correction at next-to-leading order in $\mathcal{O}(v^2)$ (excluding
the leading order result). The long distance matrix elements are
accurate up to leading order in $\alpha_s$ from Eq. (\ref{matrixlo})
or next-to-leading order in $\alpha_s$ from Eq.
(\ref{matrixenhanced}). The variable $p_t$ is set to be from 5~GeV
to 30~GeV (50~GeV) for the Tevatron (LHC), and the distributions are
depicted in logarithm unit along the vertical axis. All curves
decrease rather rapidly as the transverse momentum $p_t$ increases,
and the leading order $d\sigma/d{p_t}$ behavior is not changed by
the relativistic corrections. It can be seen that the ratio of
relativistic correction to leading order term is $1\%$ or so, and
less than $2\%$, which is
insignificant and negligible. 

The tiny effect of relativistic corrections is partly due to the
smallness of the short distance coefficient correction. In fact, the
ratio of NLO short distance coefficient to LO one from
Eq.~(\ref{G3S1}) and Eq.~(\ref{F3S1}) can be expanded as a series of
the small quantity $m_c$, as compared with $\sqrt{s}$, and this
series reduces to a fixed small number $\frac{1}{6}$ if only the
leading order term is kept, i.e.,
\begin{equation}
\frac{G({}^3S_1^{[1]})}{F({}^3S_1^{[1]})}\rightarrow
\frac{1}{6},~~~\mbox{as}~~~ \frac{2m_c}{\sqrt{s}}\rightarrow 0,
\frac{2m_c}{\sqrt{t}}\rightarrow 0.
\end{equation}
Together with the suppression from long distance matrix elements,
the tiny effect of relativistic corrections can be accounted for.
Our results for relativistic corrections in the process
$p+\bar{p}(p)\rightarrow J/\psi+X$ are similar to that in the
$J/\psi$ photoproduction process discussed in Ref.\cite{Ma:2000}.
These results may indicate that the nonrelativistic approximation in
NRQCD is good for charmonium production at high energy collisions,
and relativistic corrections are not important. This is in contrast
to the case of double charmonium production in $e^+e^-$ annihilation
at $B$ factories, where relativistic corrections may be significant.




\section{Summary}
In this paper, relativistic corrections to the color-singlet
$J/\psi$ hadroproduction at the Tevatron and LHC are calculated up
to $\mathcal{O}(v^2)$ in the framework of NRQCD factorization
approach. The perturbative short distance coefficients are obtained
by matching full QCD differential cross section with NRQCD effective
field theory calculation for the subprocess $g+g\to J/\psi+g$. The
nonperturbative long distance matrix elements are extracted from
experimental data for $J/\psi$ hadronic and leptonic decay widths up
to $\mathcal{O}(v^2)$ with an approximate relation between the
production matrix elements and decay matrix elements. Using the
CTEQ6 parton distribution functions, we then calculate the LO
production cross sections and relativistic corrections for the
process $p+\bar{p}(p)\rightarrow J/\psi+X$ at the Tevatron and LHC.
We find that the $\mathcal{O}(v^2)$ relativistic corrections to the
differential cross sections over a wide range of large transverse
momentum $p_t$ are tiny and negligible, only at a level of about
$1\%$.
The tiny effect of relativistic corrections is due to the smallness
of the short distance coefficient correction and the suppression
from long distance matrix elements. These results may indicate that
the nonrelativistic approximation in NRQCD is good for charmonium
production at high energy hadron-hadron collisions, and relativistic
corrections can not offer much help to resolve the large discrepancy
between leading order prediction and experimental data for $J/\psi$
production at the Tevatron.  Other mechanisms such as those
suggested in \cite{Haberzettl:2007kj,Lansberg:2005pc,Nayak:2007mb}
may need to be considered, aside from higher order QCD
contributions.

\section{ACKNOWLEDGEMENT}
We would like to thank Dr. Ce Meng for reading the manuscript and
helpful discussions. This work was supported by the National Natural
Science Foundation of China (No 10675003, No 10721063).


\end{document}